\documentstyle[preprint,aps,epsf]{revtex} 
\tighten
\begin{document}
\preprint{\small OUTP-96-58P/UUITP-5-97}
\draft 
\title{Multiple inflation}
\author{\large Jennifer A. Adams \bigskip}
\address{Department of Theoretical Physics, Uppsala University, \\ 
         Box 803, S-75108 Uppsala, Sweden}
\author{\large Graham G. Ross \ \& \ 
 Subir Sarkar~\footnote{PPARC Advanced Fellow\\ 
 \hspace*{2mm} hep-ph/9704286} \bigskip}
\address{Theoretical Physics, University of Oxford, \\
         1 Keble Road, Oxford OX1 3NP, U.K. \bigskip}
%\date{Submitted to {\sl Nuclear Physics B}}
\maketitle

\begin{abstract} 
Attempts at building an unified description of the strong, weak and
electromagnetic interactions usually involve several stages of
spontaneous symmetry breaking. We consider the effects of such
symmetry breaking during an era of primordial inflation in
supergravity models. In cases that these occur along flat directions
at intermediate scales there will be a succession of short bursts of
inflation which leave a distinctive signature in the spectrum of the
generated scalar density perturbation. Thus measurements of the
spectral index can directly probe the structure of unified theories at
very high energy scales. An observed feature in the power spectrum of
galaxy clustering from the APM survey may well be associated with such
structure. If so, this implies a characteristic suppression of the
secondary Doppler peaks in the angular power spectrum of temperature
fluctuations in the cosmic microwave background.
\end{abstract}
\bigskip

\widetext
\small

\section{Introduction}

Cosmological inflation is an attractive idea in search of a basis in
physics beyond the Standard Model \cite{inflbook}. It has become
increasingly clear over the past decade that the most compelling such
extension of physics is the local version of supersymmetry (SUSY),
i.e. supergravity (SUGRA) \cite {susyrev}. This provides a natural
framework for obtaining the required extremely flat and radiatively
stable potential for the scalar `inflaton' field $\phi$, the large and
approximately constant vacuum energy of which drives an exponential
increase of the scale-factor and is then converted to radiation when
the inflaton settles into its global minimum \cite
{oliverev,lythrev}. There has also been impressive progress in
astronomical observations of large-scale structure (LSS) in the
universe \cite{lss} and of angular anisotropy in the cosmic microwave
background (CMB) \cite{cmbobs}, which provide detailed constraints on
the spectrum of quantum fluctuations generated during the inflationary
De Sitter era, and therefore on the inflaton potential
\cite{inflrev}. Together these advances promise to shed light on the
structure of the theory at very high energies for it is known
\cite{inflbook,inflrev} that the amplitude of the large angular scale
CMB fluctuations observed by COBE require the energy scale $\Lambda$
of the responsible inflationary era to be well above the electroweak
scale.\footnote{\footnotesize These fluctuations are on angular scales
 much bigger than the (apparent) causal horizon at the last scattering
 surface of the CMB, thus arguing for an inflationary origin. The
 fluctuation amplitude is proportional to the Hubble rate during
 inflation, hence to $\Lambda^2$.} We refer to this era as `primordial
inflation'.

Most SUGRA models of inflation studied so far contain the inflationary
potential in a sector termed `hidden' because it is coupled only
gravitationally to the visible sector containing the quarks and
leptons and the states responsible for their non-gravitational
interactions. This is done because the required flatness of the
inflaton potential is easily maintained in the hidden sector. In such
schemes, the physics of the visible world appears to play no
significant role during inflation. In this paper we will demonstrate
that this is not necessarily the case because the visible sector may
undergo spontaneous symmetry breaking {\em during} inflation. This
can have significant effects on the quantum fluctuations of the
inflaton through the inevitable gravitational interactions.

There are two essential requirements for this to have observationally
interesting effects. The phase transition should occur during
inflation and the subsequent amount of inflation should be small
enough so as not to inflate the density perturbation generated during
the phase transition to scales beyond the present horizon. Both
conditions are naturally met in supergravity models.

To see this, let us consider the origin of symmetry breaking in
supergravity theories. We view these as the effective ``low-energy''
theories originating from the underlying unified `Theory of
Everything', the only viable candidate for which is the
superstring. In the latter, the fundamental scale is the string
tension which may be re-interpreted as the Planck scale; all other
scales must be generated by spontaneous symmetry breakdown. How does
this occur? The first requirement is a source of SUSY breaking. The
origin of this is not firmly established but the most promising
mechanism is through a stage of dynamical symmetry breaking involving
a strongly coupled hidden sector. In analogy with QCD it is then
reasonable to assume that the strongly interacting fermions --- the
gaugino partners of the gauge boson of the strong group --- form a
condensate which necessarily breaks local SUSY. The scale of SUSY
breaking is the scale at which the coupling becomes strong and thus,
like QCD, can be far from the Planck scale. Once SUSY is broken, this
breaking is communicated to all sectors of the theory via
gravitational interactions.

For our present purposes all we need to assume is {\em some} mechanism
that gives such an initial source of `soft' SUSY breaking mass terms
for all the fields of the theory. The scale for these is set by the
gravitino mass $m_{3/2}$, which must be of ${\cal O}(m_W)$ if the
hierarchy problem is to be solved allowing for a small electroweak
breaking scale. Now consider the possible sources of gauge symmetry
breaking. In supersymmetric theories there are many `flat' directions,
even amongst fields with large gauge (or other) couplings. These are
directions in field space along which, in the absence of SUSY
breaking, the potential identically vanishes or occurs only in higher
order, suppressed by some power of the Planck scale. Along such
directions, which we will characterize by the field $\rho$, the SUSY
breaking mass terms will be the dominant terms in the potential. If
the mass-squared is negative at the origin of field space, symmetry
breaking occurs along the flat direction and the vacuum expectation
value (vev) of $\rho$, which we denote by $\Sigma$, will be determined
by the stabilizing higher order terms. Since these are small, the vev
$\Sigma$ will be very large and can be close to the Planck scale. The
important point is that the SUSY breaking trigger for fields in the
visible sector is small, of ${\cal O}(m_{3/2})$. This is true even
though the scale of symmetry breaking thus generated is very large. It
is now easy to see why such a stage of symmetry breaking in the
visible sector must {\em necessarily} occur during primordial
inflation. The reason is that the gauge and/or Yukawa interactions of
the field $\rho$ will ensure that it is in thermal equilibrium before
inflation. Thus there will be stabilizing terms $\propto\,\rho^2T^2$
in the full effective potential $V(\phi,\rho,T)$ which will prevent
the transition to the asymmetric minimum from occurring until the
temperature drops to $T\sim\,m_{3/2}$. This happens only after several
e-folds of inflation as we shall demonstrate. Thereafter the phase
transition will occur but this will affect the inflaton sector only
through gravitational corrections. If these are small, inflation will
continue.

Thus we have the first ingredient necessary for imprinting structure
onto the (nearly) scale-invariant scalar density perturbation
generated during inflation because, as the phase transition occurs,
the mass of the inflaton changes through its gravitational coupling to
$\rho$. This would be irrelevant if inflation continues afterwards for
more than about $50$ e-folds for then any such structure would then
be inflated to spatial scales larger than our present horizon. However
this may not happen if there is more than one stage of intermediate
scale symmetry breaking. Then there will be several phase transitions
in succession each of which imprints structure on the density
perturbation. At each stage there is the possibility that the
gravitational coupling of the inflaton to the visible sector may
change the inflaton potential to such an extent that it causes
inflation to end. This provides the second essential ingredient in
that it makes it likely that the structure in the density perturbation
produced during the previous phase transitions will be inflated by a
{\em limited} number of e-folds and thus may still remain within our
observable horizon.

This is not yet the end of the story because on reheating after
inflation, the symmetries in the visible sector may be restored by
thermal effects.  Thus as the universe cools during the
radiation-dominated era, the field $\rho$ may {\em again} be trapped
at the origin until the temperature drops below $T\sim \,m_{3/2}$. Now
the potential energy in the visible sector will dominate and generate
a late era of inflation. When this behaviour was first noticed a
decade ago \cite{crisis,interm} it was considered to be an ``entropy
crisis'' since it would erase the baryon asymmetry presumed to have
been generated at the GUT scale. We know now that there are several
plausible mechanisms for low temperature baryogenesis well below the
GUT scale, for example electroweak baryogenesis \cite{baryogen} as
well as the Affleck-Dine mechanism which is specific to the flat
directions under discussion \cite{sugrabaryo}. Therefore late
inflation is no longer a fatal problem. Indeed the above mechanism has
recently been revived \cite {thermalinfl} under the name `thermal
inflation' to offer a plausible solution to the problem of
overproduction (after primordial inflation) of moduli fields with
electroweak scale (or heavier) masses. Again the crucial observation
is that only a limited number of e-folds of inflation occur, so such
late inflation \cite{lateinfl} does not destroy the success of
primordial inflation in providing the density perturbation responsible
for the CMB anisotropy observed by COBE and necessary to seed the
growth of the observed large-scale structure.

With this motivation we now turn to a quantitative study of such
multiple inflation in supergravity models. A preliminary account of
this work was presented earlier \cite{warsaw}.

\section{The scalar potential in supergravity}

In this section we review the form of a $N=1$ supergravity
potential. The results of relevance to the inflationary potential will
be summarised in the next section so the reader not interested in the
fine details may skip this section.

In constructing a model of inflation we are interested in the form of
the potential for the scalar `inflaton' field, {\em viz.} the order
parameter associated with a phase transformation which generates a
period of exponentially fast growth of the cosmological scale
parameter. In $N=1$ supersymmetric theories with a single SUSY
generator, complex scalar fields are the lowest components, $\phi^a$,
of chiral superfields, $\Phi^a$, which contain chiral fermions,
$\psi^a$, as their other components. In what follows we will take
$\Phi^a$ to be left-handed chiral superfields so that $\psi^a$ are
left-handed massless fermions. Masses for fields will be generated by
spontaneous symmetry breakdown so that the only fundamental mass scale
is the normalized Planck scale, $M\equiv\,M_{\rm
P}/\sqrt{8\pi}\simeq2.44\times10^{18}$~GeV. This is aesthetically
attractive and is also what follows if the underlying theory
generating the effective low-energy supergravity theory follows from
the superstring. The $N=1$ supergravity theory describing the
interaction of the chiral superfields is specified by the K\"{a}hler
potential \cite{susyrev},
\begin{equation}
 G (\Phi, \Phi^\dagger) = d (\Phi, \Phi^\dagger) + \ln |f (\Phi)|^2 .
\label{g}
\end{equation}
Here $d$ and $f$ (the superpotential) are two functions needed to specify
the theory; they must be chosen to be invariant under the symmetries of the
theory. The dimension of $d$ is 2 and that of $f$ is 3, so terms bilinear
(trilinear) in the superfields appear without any mass factors in $d$ ($f$).
The scalar potential following from eq.(\ref{g}) is given by \cite{susyrev} 
\begin{equation}
 V = \exp \left(\frac{d}{M^2}\right) \left[F^{A^\dagger} (d_A^B)^{-1} F_B 
      - 3 \frac{|f|^2}{M^2} \right] + {\rm D-terms}\ ,  
\label{V}
\end{equation}
where 
\begin{equation}
 F_A \equiv \frac{\partial f}{\partial \Phi^A} 
     + \left(\frac{\partial d}{\partial \Phi^A}\right) \frac{f}{M^2}\ , \qquad 
 \left(d_A^B\right)^{-1} \equiv \left(
  \frac{\partial^2 d}{\partial \Phi^A \partial \Phi_B^\dagger} \right)^{-1} .
\end{equation}
We follow ref.\cite{sw} in expanding $f$ and $d$ as a series in $M$ 
\begin{eqnarray}
 f (z, y) &=& M^2 f^{(2)} (\case{z}{M}) + M f^{(1)} (\case{z}{M}) 
  + f^{(0)} (\case{z}{M}, y)\ ,  \nonumber \\
 d (z, z^\dagger, y, y^\dagger) &=& M^2 d^{(2)} (\case{z}{M}, 
  \case{z^\dagger}{M}) + M d^{(1)} (\case{z}{M}, \case{z^\dagger}{M}) 
  + d^{(0)} (\case{z}{M}, \case{z^\dagger}{M}, y, y^\dagger)\ .
\label{fd}
\end{eqnarray}
Here $z$ are chiral superfields whose scalar component have large vevs
and $y$ correspond to those fields which do not have large vevs and
which remain light. Thus the superfields $y$ cannot appear multiplied
by powers of $M$, hence the form of eq.(\ref{fd}).

If, as is the case in theories descending from the superstring, there
are no mass scales other than $M$ and those induced by spontaneous
breaking, we have for the terms generating the renormalizable self
couplings of the light fields $y$ the form \cite{gm}
\begin{eqnarray}
 f^{(0)} (\case{z}{M}, y) &=& \sum_n c_n (z) g_n^{(3)} (y)\ , \nonumber \\
 d^{(0)} (\case{z}{M}, \case{z^\dagger}{M}, y, y^\dagger) &=& 
  y^a \Lambda_a^b (\case{z}{M}, \case{z^\dagger}{M}) y_b^\dagger 
  + \sum_m c^{\prime}_m (\case{z}{M}, \case{z^\dagger}{M}) g_m^{(2)}(y) 
  + {\rm h.c.}\ ,
\label{fd0}
\end{eqnarray}
where $g_n^{(3)}(y)$ and $g_m^{(2)}(y)$ are, respectively, the
trilinear and bilinear terms in $y^a$ allowed by the symmetries of the
theory. In addition there will be further terms in eq.(\ref{fd0})
suppressed by inverse powers of $M$ which generate non-renormalizable
terms involving the light fields $y$ in the effective low energy
Lagrangian.

From ref.\cite{gm}, the terms in the effective potential which
survive in the $M\rightarrow\infty$ limit (keeping terms of order
$z/M$) are:
\begin{eqnarray}
 V &\approx& \left|\frac{\partial \hat{g}}{\partial y^a}\right|^2 
   + m_{3/2}^2 y^a S_a^b y_b^\dagger 
   + [m_{3/2}^\dagger (y^a R_a^b \frac{\partial \hat{g}}{\partial y^b} 
   + \sum_n (A_n - 3) \hat{g}^{(3)}_n  \nonumber \\
&& + \sum_n (B_m - 2) \mu_m g_m^{(2)}) + {\rm h.c.}] + {\rm D-terms}\ ,
\label{va}
\end{eqnarray}
where 
\begin{equation}
 m_{3/2} \equiv \langle{\rm e}^{d^{(2)}/2} f^{(2)}\rangle  
\label{mg}
\end{equation}
is the gravitino mass, and 
\begin{equation}
 S_a^b = \delta^b_a + M^2 \left\langle\rho^{\dagger i}\left(\frac{\partial
         \Lambda_c^b} {\partial z^\dagger_j} \frac{\partial \Lambda_a^c} 
         {\partial z^i} - \frac{\partial^2 \Lambda_a^b} 
         {\partial z^\dagger_j \partial z^i} \right)\rho_j\right\rangle , 
 \qquad 
 R_a^b = \delta_a^b - M \left\langle\rho^{\dagger i} 
  \frac{\partial \Lambda_a^b}{\partial z^i} \right\rangle ,
\label{sr}
\end{equation}
where
\begin{equation}
 \rho_j \equiv \left(M \frac{\partial^2 d^{(2)}} {\partial z^i \partial
  z^\dagger_j}\right)^{-1} 
 \frac{\partial}{\partial z^j}(\ln f^{(2)}+d^{(2)})\ .
\label{rho}
\end{equation}
Here $\hat{g}$ is the superpotential for the light fields defined by
\begin{equation}
 \hat{g}(y) = \sum_n \hat{g}^{(3)}_n (y) + \sum_m \mu_m g_m^{(2)} (y)\ ,
\end{equation}
with 
\begin{equation}
 \hat{g}^{(3)}_n (y) = \langle {\rm e}^{d^{(2)}/2} S \rangle 
  c_n (\langle{\case{z}{M}}\rangle) g_n^{(3)} (y)\ , \qquad 
 \mu_m = m_{3/2} \left\langle 
  \left(1 - M\rho_i\frac{\partial}{\partial z^\dagger_i}\right)
  c_m^{\prime}({\case{z}{M}}, {\case{z^\dagger}{M}}) \right\rangle .
\end{equation}
The coefficients $A_n$ and $B_m$ are given by 
\begin{eqnarray}
 A_n &=& M \left\langle \rho^{\dagger i} \frac{\partial}{\partial z^i}  
         [d^{(2)} + \ln c_n (\case{z}{M})] \right\rangle , \nonumber\\ 
 B_m &=& \left\langle \left[2 + M \left(
           \rho^{\dagger i} \frac{\partial}{\partial z^i}
         - \rho_i \frac{\partial}{\partial z^\dagger_j}\right) 
         - M^2 \rho^{\dagger i} \rho_j \frac{\partial^2}{\partial z^i 
            \partial z^\dagger_j}\right] 
         \frac{c'_m (\case{z}{M}, \case{z^\dagger}{M})} 
          {(1 - M \rho_i \frac{\partial} 
          {\partial z^\dagger_i} c'_m 
          (\case{z}{M}, \case{z^\dagger}{M})} \right\rangle .
\end{eqnarray}

\subsection{Constraints on the inflaton potential}

The question whether the form of eq.(\ref{va}) is such as to generate
a period of inflation has been considered extensively in the
literature \cite {lythrev}. For our purposes here it is sufficient to
introduce a general formalism encompassing the various possibilities
by expanding the (slowly varying) potential about the value $\phi
^{*}$ in inflaton field space at which the quantum fluctuations of
interest are produced. Thus we write $\phi =\tilde{\phi}+\phi ^{*}$
(in units of $M$) and expand the inflaton potential as
\begin{equation}
 V_{\rm I}(\phi) = \Lambda^4\left[1 
  + c_{1}\tilde{\phi}
  + c_{2}\tilde{\phi}^2
  + c_{3}\tilde{\phi}^3
  + c_{4}\tilde{\phi}^4 + \dots \right]\ .
\label{expand}
\end{equation}
Here we have factored out the overall scale of inflation $\Lambda $
which is usually required to be $\sim10^{-4}M$ in order to account for
the CMB anisotropy observed by COBE
\cite{sic,success}.\footnote{\footnotesize Later we will discuss ways
in which this constraint may be relaxed.} Note that since a non-zero
potential breaks supersymmetry, the magnitude of SUSY breaking during
inflation is {\em different} from that after inflation. During
inflation the gravitino mass receives a contribution of order the
Hubble parameter \cite {susybreak} which, in this case, is
\begin{equation}
 |m_{3/2}| \sim \frac{\Lambda^2}{M}\sim 10^{-8}M\ ,  
\label{mf}
\end{equation}
i.e. much larger than its nominal value of ${\cal O}(m_{W})$.

The constraints on the parameters in the potential following from the
`slow-roll' conditions \cite{inflrev} for successful inflation are
\begin{eqnarray}
 c_1 \ll 1\ , \quad c_2 \ll 1\ , \quad c_3 \tilde{\phi} \ll 1\ , 
 \quad c_4 \tilde{\phi}^2 \ll 1\ ,\ \ldots  
\label{bound}
\end{eqnarray}
We focus on `new inflation' in which initially $\tilde\phi\ll\,1$, so
the constraints are only severe for $c_1$ and $c_2$. A variety of
models have been examined in which these constraints are automatically
satisfied \cite {lythrev}. For example symmetry may give $c_1,~c_2=0$
while $c_3$ and $c_4$ are of ${\cal O}(1)$ \cite{sic,success}. More
interestingly, the problematic quadratic term may dynamically relax to
zero by virtue of an infrared fixed point at the origin for the
coupled equations of motion of the inflaton field $\tilde{\phi}$ and
the modulus field $\varphi$ which determines its couplings
\cite{natural}. This requires only that the kinetic term have a
symmetry leading to (pseudo) Goldstone modes and the resulting scalar
potential then has the form (\ref{expand}) with the coefficients
\begin{equation}
 c_1 = 0\ , \quad  
 c_2 \sim \varphi\ , \quad 
 c_3 \sim 1\ , \quad 
 c_4 \sim 1\ ,\ \ldots
\label{vf}
\end{equation}
The fields are driven to the initial values
$\varphi,|\tilde{\phi}|\sim\,H$ so the `quadratic' and cubic terms
become comparable; thereafter $\tilde{\phi}$ grows more rapidly than
$\varphi$ and the cubic term soon dominates \cite{natural}. This
results in a `tilted' spectrum for the scalar density perturbation
which solves the problem of excess power on small spatial scales in
the COBE-normalized cold dark matter (CDM) cosmogony.

For the present discussion, the precise form chosen is not crucial, so
we proceed under the assumption that there {\em is} a hidden sector
inflationary potential satisfying the slow-roll conditions.

\section{Multiple primordial inflation}

As discussed above, supersymmetric theories may have `flat' directions
in field space along which combinations of fields may acquire vevs
without changing the potential significantly. Gauge singlet moduli
fields of the type found in superstrings \cite{strings} provide an
example of this, having absolutely flat potentials in the absence of
SUSY breaking. The same may be true for strongly interacting fields
e.g. gauge non-singlets may acquire vevs along `D-flat'
directions. Along these flat directions it is the SUSY breaking masses
that determine the vevs. For example, if the potential for a field $y$
is {\it completely} flat in the absence of SUSY breaking, the
superpotential $f$ in eq.(\ref{g}) vanishes identically and the scalar
potential is given by eq.(\ref{va}) as
\begin{equation}
 V \approx m_{3/2}^2 (S|y|^2 + Ty^2 + {\rm h.c.})\ ,
\label{vinf}
\end{equation}
where 
\begin{equation}
 S_{b}^{a} = S\delta _{b}^{a}\ , \qquad 
 T = (B_{m} - 2 + 2R) 
  \left\langle\left(1 - M \rho_i\frac{\partial}{\partial z_i^\dagger}\right) 
  c_{m}^\prime ({\case{z}{M}}, \case{z^\dagger}{M})\right\rangle 
  \frac{\partial^2 g_{m}^{(2)}}{\partial y^2}\ , \qquad 
 R_{b}^{a}=R\delta_{b}^{a}\ .
\label{ST}
\end{equation}
We see that the scalar potential is proportional to the square of the
SUSY breaking parameter $m_{3/2}$, thus vanishing in the
supersymmetric limit.

Let us now consider the potential for a field $\rho $ which triggers a
stage of symmetry breaking in the visible sector (for simplicity $\rho
$ may be taken to be real). The full scalar potential describing the
inflaton field $\phi $ and the visible sector field $\rho $ is
\begin{equation}
 V = V_{\rm I}(\phi) + m^2(\case{\rho}{M})\ 
  \rho^2 [1 + {\cal O}(\case{\rho}{M})]\ .
\label{fp}
\end{equation}
Here we have allowed for the effect of radiative corrections in the
visible sector which cause the SUSY breaking mass $m$ to run, with
corrections dependent on the logarithm of $\rho/M$. If $m^2$ is
positive at the Planck scale but changes sign due to these radiative
corrections at a scale $\Sigma$, the potential will have a minimum
at an `intermediate' scale very close to $\Sigma$. In the case the
flat direction described by $\rho$ is only approximately flat, there
will be additional terms in the potential proportional to $\rho
^{N}/M^{N-4}$ which will provide a stable minimum even if the SUSY
breaking mass-squared is negative at the Planck scale. However even in
this case the vev of the field $\rho$ at the minimum will be large,
of ${\cal O}(M^{N-4}m_{3/2}^2)^{1/(N-2)}$. Let us denote this scale
too by $\Sigma $. For future reference we note that the depth of the
potential in both cases is given approximately by
$m_{3/2}^2\Sigma^2\sim\Lambda^4(\Sigma/M)^2$.

We are interested in the case $\rho$ is a visible sector field having
strong (gauge and/or Yukawa) interactions which will maintain it
initially in thermal equilibrium with a potential of the form
\cite{crisis}
\begin{equation}
 V (\phi ,\rho , T) = V_{\rm I} (\phi) + 
  \left[-|m|^2(\case{\rho }{M}) + \alpha T^2\right] \rho^2
  [1 + {\cal O}(\case{\rho}{M})]\ .
\label{is}
\end{equation}
where we have now explicitly indicated that $m^2$ is negative. This
form applies for values of the $\rho$ field small enough that its mass
is less than the temperature. Since $\rho$ has strong couplings this
essentially requires $\rho<T$. Above this scale the thermal term in
the effective potential above vanishes. Thus the thermal term creates
a potential barrier of height $\sim\,T^4$ in between $\rho\sim\,T$ and
$\rho\sim\,T^2/|m|$, which prevents the vev of $\rho$ from developing
until the temperature drops to $T_{\rm f}\sim\,|m|\approx\,|m_{3/2}|$
and the barrier disappears \cite{crisis,aspects}. However, before this
happens but after $T$ drops below $T_{\rm
i}\sim\,V^{1/4}\approx\Lambda$, the approximately constant potential
term $V_{\rm I}(\phi)$ dominates the thermal energy, driving a short
period of inflation. Note that $T_{\rm f}$ is determined by the value
of $m$ (given by eq.(\ref{mf})) {\em during} this period of inflation,
i.e. $T_{\rm f}\approx\,V_{\rm I}^{1/2}/M$. The important point is
that the extent of this inflationary period is thus limited to
$\ln(T_{\rm i}/T_{\rm f})\sim\ln(M/\Lambda)$ i.e. about 10 e-folds for
$\Lambda\sim10^{-4}M$. Subsequently the field $\rho$ evolves to its
global minimum at $\Sigma$ according to the governing equation
\begin{equation}
 \ddot{\rho} + 3H \dot{\rho} + \frac{{\rm d}V}{{\rm d}\rho} = 0\ ,
\label{evolve}
\end{equation}
so its growth (for $\langle \rho \rangle \ll \Sigma $) obeys 
\begin{equation}
\rho =\rho _{0}\exp \left[ \frac{3Ht}{2}\left( \sqrt{1+\frac{8m^2}{9H^2}}
-1\right) \right] .
\end{equation}
Since $|m|\sim \,H$ we have $\rho \sim \rho _{0}{\rm e}^{Ht}$ with
$\rho _{0}\lesssim\,T_{\rm f}$, so there is a further
$\sim\ln(\Sigma\,M/\Lambda^2)$ e-folds of inflation before $\rho$
settles into its minimum and releases its stored energy. (However,
because of the exponentially fast growth, most of the actual increase
in $\langle\rho\rangle$ occurs in the last 1--2 e-folds.) Thus the
total number of e-folds of inflation is:
\begin{equation}
 {\cal N}_{1}^{\rm prim} \sim 
  \ln \left( \frac{\Sigma M^2}{\Lambda^3}\right) .
\label{tot}
\end{equation}
As the intermediate scale transition occurs energy is released and
there is the possibility of reheating in the visible sector to which
$\rho$ is coupled. However fields strongly coupled to $\rho$ will
obtain a large intermediate scale mass and thus will not be readily
produced by $\rho$ decay or by its rescattering products. Thus one
must consider only reheating processes to visible sector states $\chi$
which have coupling $h\rho^2\chi^2$ with the coupling constant $h$
constrained by $h\langle\rho\rangle<T_{\rm R}$, where $T_{\rm R}$ is
the reheat temperature. With such a small coupling the energy release
does not lead to reheating above the Hawking temperature $T_{\rm
H}=H/2\pi$ in the De Sitter vacuum because of the large Hubble
parameter during primordial inflation.

After the phase transition inflation continues but with a reduced
scale, i.e. 
\begin{equation}
  V_{\rm I} \rightarrow 
   V_{\rm I} \left[1 - \left(\frac{\Sigma}{M}\right)^2\right] .
\end{equation}
This will have the effect of changing the amplitude of the generated
density perturbation (see eq.(\ref{deltah})), albeit by a negligible
amount for small $\Sigma /M$. A more significant change to the density
perturbation will occur however due to the inevitable gravitational
coupling between the $\rho$ and $\phi$ sectors, changing the
inflaton mass. Note that since the latter is anomalously small
relative to the Planck scale, small changes in it have a
disproportionately large effect. The magnitude of this effect is
easily estimated by adding to the effective low energy theory, terms
allowed by the governing symmetries. Thus we may expect
\footnote{\footnotesize Here we have assumed that $\phi$ transforms
 non-trivially under the symmetries of the theory so that
 $\phi\phi^\dagger$ is the first invariant. We also assume that
 inflation occurs near $\phi=0$, i.e. $\phi^*=0$ in
 eq.(\ref{expand}). It is easy to generalize the discussion to cover
 other possibilities.} in eq.(\ref{g}) a term
$d\supset\beta\phi\phi^\dagger\rho^2/M^2$. Following from the scalar
potential (\ref{V}) this gives a contribution to the inflaton mass of
\footnote{\footnotesize From eq.(\ref{V}) it is easy to see that the
 change in V$_{\rm I}$ due to the intermediate scale transition also induces
 changes in $m_\phi$ of the same order.}
\begin{equation}
 \delta m_{\phi }^2 \sim m_{3/2}^2\ \beta \left(\frac{\Sigma}{M}\right)^2 ,
\label{cim}
\end{equation}
to be compared with the effective inflaton mass during inflation given by
eqs.(\ref{expand}) and (\ref{vf}),
\begin{equation}
 m_{\phi }^2 = 
  m_{3/2}^2\ c_{3} \left(\frac{\langle|\tilde{\phi}|\rangle}{M}\right) .
\end{equation}
For $\beta(\Sigma /M)^2$ of ${\cal
O}(c_{3}(\langle|\tilde{\phi}|\rangle/M))$, the inflatonary potential
will be significantly modified and hence so too will be the density
perturbation which is sensitive to the inflaton mass.

Of course if the second period of inflation continues for more than
about $50$ e-folds the difference in the density perturbations
produced in the two periods of inflation becomes observationally
irrelevant. However, analyses of realistic models suggest there may be
several flat directions and several stages of spontaneous symmetry
breaking, so we may expect this new period of inflation to be
short-lived as well because of a {\em further} stage of intermediate
scale symmetry breaking (at a scale $\Sigma_2$) either running
concurrently or triggered by the first stage of breaking (through the
effect on the mass of the second `flat direction' field of the vev of
the first). This second period of inflation lasts just
\begin{equation}
 {\cal N}_{2}^{\rm prim} \approx 
  \ln\left(\frac{\Sigma_2}{\Sigma}\right) ,\ 
  \ln \left(\frac{\Sigma_2 M}{\Lambda^2}\right) 
\end{equation}
e-folds for the two cases respectively. As before, after the second
stage of breaking is completed, the form of the inflation potential
may change significantly due to the gravitational coupling between the
inflaton and `flat' sectors. The whole process may continue for as
many stages of intermediate scale breaking as there are flat
directions.

What about the end of inflation? There are two possibilities. The
first is that the end occurs in the usual manner when the inflaton vev
becomes of ${\cal O}(M)$. Since the overall number of e-folds of
inflation in a `slow-roll' model is very large (see
e.g. \cite{success,natural}), one may expect that the total number of
e-folds in the last period of inflation, after intermediate scale
breaking, is also quite large, much greater than $50$, so the effect
of the previous eras of intermediate scale breaking is observationally
irrelevant. The second, more interesting, possibility is that at the
end of one of the intermediate breaking stages the modification of the
inflaton potential is so great that it ceases to satisfy the slow-roll
conditions. As we have discussed ({\em c.f.}  eq.(\ref{cim})) this
happens if $\beta(\Sigma_i/M)^2\sim1$, for then the change in the
$\phi$ potential violates eq.(\ref{bound}). In this case the
corresponding intermediate scale phase transition {\em ends}
inflation. Note that this may occur at any of the symmetry breaking
stages since there is no requirement for the various intermediate
scales $\Sigma _i$ to be all of the same magnitude. This follows
because these depend sensitively on the form of the non-renormalizable
higher dimension terms which lift the flatness of the potential in the
supersymmetric limit.

When inflation is terminated, the potential energy $\Lambda^4$ of the
inflaton $\phi$ is released. Reheating is inefficient in our model
\cite{sic} as the inflaton has only gravitational couplings to the
visible sector resulting in a slow decay rate
$\Gamma_{\phi}\sim\Lambda^6/M^5$. As discussed earlier \cite{success},
the temperature at the beginning of the standard radiation dominated
era is consequently low:
\begin{equation}
 T_{\rm R}^{\rm prim} \sim 
  \left(\sqrt{\frac{30}{\pi^2 g_\ast}} \Gamma_{\phi} M \right)^{1/2} \sim 
  10^5\ {\rm GeV} ,
\label{Trprim}
\end{equation}
taking $g_\ast$, the effective number of relativistic degrees of
freedom in the plasma, to be 915/4 as in the MSSM for $T\gg\,m_{W}$.
There is no parametric resonance \cite{reheat} since the inflaton has
no coupling to the visible sector fields $\chi$ of the form
$\phi^2\chi^2$ but only terms involving $\chi^3$ \cite{success}. This
is because the supergravity couplings involve the square of the
superpotential $f$ which is {\em trilinear} in the scalar fields,
therefore bilinear terms are suppressed by $m_{\chi}/M$. Note that
even if the reheating were to be prompt due to an enhanced decay rate,
energy conservation imposes the absolute bound
$$
  T_{\rm R}^{\rm prim} 
   \sim \left(\sqrt{\frac{30}{\pi^2 g_\ast}} H M \right)^{1/2}
   \lesssim \frac{\Lambda}{3}\ .
$$

\subsection{The scalar density perturbation}

The case inflation ends due to an intermediate scale transition after
a series of such intermediate scale transitions implies a novel and
interesting structure for the spectrum of the scalar density
perturbation.  In this picture our observational universe has resulted
from a succession of inflationary periods, each lasting a relatively
small number e-folds. The magnitude of the density perturbation during
each period is determined by a combination of the parameters in the
inflationary sector and in the intermediate scale sector. Thus we have
a natural model for features in the perturbation spectrum which may be
probed by observations of LSS and the CMB.

The magnitude of the density perturbation produced in a theory with
multiple scalar fields has been discussed in ref.\cite{multi}
following the formalism of ref.\cite{pert}.\footnote{\footnotesize
 Strictly speaking this discussion applies in the slow-roll
 approximation (albeit to second-order) which is not always valid
 during the evolution of $\rho$; however the general features used here
 do apply.} During inflation, fluctuations in both the $\rho$ and
$\phi$ fields lead to density perturbations. Their impact depends on
the sensitivity of ${\cal N}$ --- the number of e-folds of inflation
from the time the fluctuations are produced until the end of inflation
--- to $\rho$ and $\phi$. We are interested in the density
perturbation produced about 40--50 e-folds before the end of inflation
since this is the range to which observations are sensitive. As we
have discussed an intermediate scale phase transition occurring at
this time will be completed well before inflation ends.  Thus the
dominant density perturbation arises from the usual quantum
fluctuations $\delta\phi\sim\,H.$ We conclude that, even though this
is a two-scalar field problem, the usual form \cite{inflrev} for the
spectrum of scalar adiabatic density perturbation {\em does} apply:
\begin{equation}
 \delta^2_{\rm H} (k) = \frac{H^2}{2\pi\dot{\phi}} 
  = \frac{1}{75\pi^2M^6} \left(\frac{V^3}{V'^2}\right)_\star ,
\label{deltah}
\end{equation}
where $\star$ denotes the epoch at which a scale of wavenumber $k$
crosses the Hubble radius $H^{-1}$ during inflation. The spectral
index, defined as below, is then given by \cite{inflrev}
\begin{equation}
 n_{\cal R}(k) - 1 \equiv \frac{{\rm d}\ln \delta^2_{\rm H}(k)}{{\rm d}\ln k}
 = - 3 M^2 \left(\frac{V'}{V}\right)^2_\star 
   + 2 M^2 \left(\frac{V''}{V}\right)_\star .
\label{n}
\end{equation}
However this does not mean that the produced density perturbation is
insensitive to the intermediate scale because the $\phi$ potential is
altered after the intermediate scale transition changing $\delta_{\rm
H}$ and $n_{\cal R}$. The magnitude of these changes depends on the
details of the theory ({\em c.f.} the dependence in eq.(\ref{cim}) on
the parameter $\beta$). Before and after the intermediate scale
transition the density perturbations are nearly scale invariant (with
a small `tilt' \cite{natural}) but the amplitude of the perturbation
may differ substantially. Inflation continues in the period between
these two bursts of inflation but the spectral index will necessarily
change from unity in order to accommodate the change in the magnitude
of the perturbation before and after the transition. Typically
$n_{\cal R}$ will drop to around zero if the terms on the rhs of
eq.(\ref{n}) become comparable. (To track $n_{\cal R}(k)$ precisely
through the phase transition requires numerical solution of the
equation of motion for the inflationary field perturbation on
spatially flat hypersurfaces \cite{pert}, as has been done for some
toy models \cite{toy}.) This change occurs in the time it takes for
the $\rho$ field to change by ${\cal O}(\Sigma )$, i.e. just 1--2
e-folds.\footnote{\footnotesize Although the $\rho$ field may take
 many more e-folds to flow to its minimum ({\em c.f.}  the discussion
 preceding eq.(\ref{tot})), over most of this period its vev changes by
 a negligible amount compared to $\Sigma$ and so does not significantly
 alter the $\phi$ potential.}

Multiple inflation also has an interesting implication for the scale
of inflation $\Lambda$ in eq.(\ref{expand}). In the case of single
inflation this scale is constrained by normalizing the generated
density perturbation (\ref{deltah}) to the COBE measurement of the CMB
anisotropy. This refers to a length scale which crossed the Hubble
radius about 50 e-folds before the end of inflation (see
eq.(\ref{Nstar})) so the corresponding value of $\phi_\star$ is
obtained by integrating the evolution equation for $\phi$ (similar to
eq.(\ref{evolve})) 50 e-folds back from the end of inflation. This is
usually assumed to occur when the slow-roll conditions are violated
due to the increasing steepening of the potential as
$\langle\phi\rangle$ approaches $M$ (e.g. ref.\cite{success}). However in
multiple inflation, the end of inflation is determined instead by an
intermediate scale transition so $\phi_{\star}$ can be significantly
{\em smaller}. As a result, $\Lambda$ may be reduced to compensate for
the reduction in $V'$ (see eq.(\ref{deltah})), while maintaining the
same amplitude for the density perturbation. Such a change would allow
us to identify $\Lambda$ with the usual SUSY breaking scale in a
hidden sector. This avoids the necessity of introducing two separate
scales \cite{thomas} or of arguing for a different scale of SUSY
breaking during and after inflation \cite{success}.

To summarize this section, we have examined the implications of an era
of primordial inflation due to a hidden sector for symmetry breaking
phase transitions in the visible sector. Transitions along flat
directions {\em necessarily} occur during inflation and can lead
naturally to a `graceful exit' from inflation. As a result primordial
inflation may occur in several short bursts producing near
scale-invariant density perturbations during each burst but with
different amplitudes. Moreover in the relatively brief period (1--2
e-folds) corresponding to the transition between the bursts,
scale-invariance is badly broken and the spectral index of the density
perturbation changes from its usual value around unity. In such a
picture the implications of the observed large-scale structure in the
universe for the cosmological parameters ($\Omega$, $\Omega_{\rm
N}$, $h$ \ldots) and the question whether a component of hot dark
matter is required must be re-examined. We discuss this in a later
section but first we consider yet another twist to the story, namely
the possibility that as primordial inflation ends, the intermediate
scale symmetry breaking fields are reheated and the symmetry is
restored. In this case the subsequent cosmological history will
involve `thermal inflation' \cite{thermalinfl}.

\section{Late thermal inflation}

After primordial inflation ends the potential energy stored in the
inflaton is released reheating the visible sector to the temperature
(\ref{Trprim}) and allowing the possibility of symmetry restoration
through thermal effects.\footnote{\footnotesize There cannot be
 symmetry restoration by non-thermal effects during the `preheating'
 phase \cite{kls}, since there is no parametric resonance in this case
 (see comment after eq.\ref{Trprim})).} This is most likely along flat
directions because the potential barrier between the symmetric and
broken symmetry phases is smallest along these directions. The
potential is again given by eq.(\ref{is}) where $V_{\rm I}$ is
constant as $\phi$ now sits at its minimum at $M$. (Note that the
scale of $m$ is now the scale of SUSY breaking {\em after} primordial
inflation, i.e. of ${\cal O}(m_W)$.) If the reheat temperature $T_{\rm
R}^{\rm prim}$ following primordial inflation (\ref{Trprim}) is
greater than $m(\Sigma/M)$, one sees that the true minimum lies at the
origin. However there is a potential barrier between this and the
minimum at $\Sigma$, with height $\sim\,m^2\Sigma^2$. Therefore only
if the reheat temperature exceeds $\sim\sqrt{m\Sigma}$ can the
symmetric minimum be reached. Note also that use of the effective
potential (\ref{is}) requires that thermal equilibrium be established
and this will {\em not} happen if the states to which $\rho$ strongly
couples have masses higher than the reheat temperature. Since these
masses are of ${\cal O}(\Sigma)$ the stronger constraint $T_{\rm
R}^{\rm prim}>\Sigma$ applies. Even allowing for the possibility that
reheating may be more efficient than the estimate in eq.(
\ref{Trprim}), the maximum possible value of $T_{\rm R}^{\rm prim}$ is
just $\Lambda$ so the condition $\Sigma<\Lambda$ is {\em essential}
for symmetry restoration. In general reheating by a hidden sector
inflaton is likely to be inefficient \cite{success} so it is a model
dependent question whether the condition necessary for symmetry
restoration will be satisfied.

If it does, intermediate scale symmetry breaking will be wiped out and
the symmetry breaking fields returned to their high temperature minima
at the origin. Thereafter the symmetry breaking will repeat itself in
a manner mirroring what happens during the era of primordial
inflation.\footnote{\footnotesize It is also possible to have
 additional stages of thermal inflation corresponding to intermediate
 phase transitions which do {\em not} occur during primordial
 inflation.}  Initially the field $\rho $ is trapped at the origin and
the potential is given by the constant $V_{\rm I}(\Sigma)$. This has
magnitude $V_{\rm I}(\Sigma)\sim\,m^2\Sigma^2$ because the present day
vacuum which {\em does} have intermediate scale breaking must have
zero cosmological constant so the change in potential energy due to
the intermediate scale breaking must just cancel its initial value. As
the temperature drops below $T_{\rm i}\sim\sqrt{m\Sigma}$ the
constant potential energy drives a further period of inflation which
continues until the temperature in the $\rho $ sector drops below
$T_{\rm f}\sim\,m$, i.e. for
\begin{equation}
 {\cal N}^{\rm therm} \simeq \ln\left(\frac{T_{\rm i}}{T_{\rm f}}\right)
  \simeq \frac{1}{2}\ln\left(\frac{\Sigma}{m_{W}}\right) ,
\end{equation}
e-folds \cite{thermalinfl}. All this happens at a low scale for the
value of $V_{\rm I}$ is now $m_{W}^2\Sigma^2\sim\Lambda _{\rm
SUSY}^4\,\Sigma^2/M^2$, where $\Lambda_{\rm
SUSY}\sim\sqrt{m_{W}M}\sim10^{10}$~GeV is the scale of SUSY breaking
in a hidden sector.

Now the $\rho$ decay rate is 
\begin{equation}
 \Gamma_\rho \sim \frac{m^3}{\Sigma^2}\ ,  
\label{rate}
\end{equation}
given a coupling $\lambda\rho\bar{\chi}\chi$ to light fields and
allowing for ${\cal O}(100)$ degrees of freedom in the final
state. Provided the fields $\chi$ to which $\rho$ couples strongly
have masses smaller than (half) its own mass and the reheat
temperature, the reheating is rapid and the potential energy is almost
entirely converted into thermal energy. The condition for this is just
$\Gamma_\rho\gtrsim\,H\approx\,m\Sigma/M$, i.e.
\begin{equation}
 \Sigma \lesssim 
  \frac{\Lambda_{\rm SUSY}^{4/3}}{M^{1/3}} \equiv \Sigma_{\rm crit}
  \approx 2\times10^7\ {\rm GeV}\ .
\end{equation}
Then the reheat temperature is given by 
\begin{equation}
 T_{\rm R}^{\rm therm} \sim \left(\frac{30}{\pi^2 g_\ast}
  m^2\Sigma^2\right)^{1/4}
 \frac{\Lambda_{\rm SUSY}}{3} \left(\frac{\Sigma}{M}\right)^{1/2} 
 \approx 10^4\ {\rm GeV} \left(\frac{\Sigma}{\Sigma_{\rm crit}}\right)^{1/2} 
 \quad {\rm for} \quad 
 \Sigma \lesssim \Sigma_{\rm crit}\ ,
\end{equation}
taking $g_\ast$, the effective number of relativistic degrees of
freedom in the plasma to be 915/4 as in the MSSM for $T\gg\,m_{W}$. If
reheating is inefficient, the value is instead (see eq.(\ref{Trprim}))
\begin{equation}
 T_{\rm R}^{\rm therm} \sim \left(\frac{30}{\pi^2 g_\ast}\right)^{1/4}
  \frac{\Lambda_{\rm SUSY}^3}{\Sigma M}
  \approx 10^4\ {\rm GeV} \left(\frac{\Sigma_{\rm crit}}{\Sigma}\right) 
 \quad {\rm for} \quad \Sigma \gtrsim \Sigma_{\rm crit}\ .
\end{equation}
The above estimates are not altered significantly even if there is a
`parametric resonance' \cite{thermalinfl,aspects}.

As mentioned earlier, the realization that baryogenesis may occur at
relatively low temperatures means that such a late stage of entropy
release is now deemed acceptable. This possibility has been promoted
recently \cite{thermalinfl} since inflation at a low scale solves the
cosmological problems associated with light fields which are generic
in supersymmetric and superstring theories \cite{lateinfl}. These are:
\begin{itemize}

\item {The `gravitino problem' associated with the generation of
massive, unstable electroweak scale particles after inflation,
e.g. the thermal production during reheating of gravitinos
\cite{gravprob}, whose decays can disrupt primordial nucleosynthesis
\cite{bbnrev}, and}

\item {The `Polonyi problem' \cite{polonyi} which refers to the
excitation during inflation of light scalar fields with flat
potentials, in particular moduli in superstring theories
\cite{moduli}. These release their stored energy rather late, thus
generating unacceptably large amounts of entropy which prevents
successful nucleosynthesis.}

\end{itemize}
Both these problems involve the production after inflation of states
with electroweak scale masses. If stable, they come to dominate the
energy density of the universe and if unstable they decay after
nucleosynthesis leading to unacceptable changes in the synthesised
elemental abundances \cite{bbnrev}.

Thermal inflation solves the gravitino problem by simply diluting to
acceptable levels the abundance of such light states produced after
primordial inflation. There is no danger of their being recreated
again afterwards as the reheat temperature following thermal inflation
is quite low. There is also the danger of the light moduli states
being produced after inflation through the conversion of potential
energy in the moduli sector released when the fields flow to their
true minimum (a.k.a. the Polonyi problem). Thermal inflation avoids
this problem as well because the potential energy driving thermal
inflation is only $m_{3/2}^2\Sigma^2$ and, through gravitational
corrections, generates a mass shift to the light field of ${\cal
O}(m_{3/2}\Sigma /M)$. This is at most comparable to the mass of
${\cal O}(m_{3/2})$ generated by the SUSY breaking sector.
Consequently the potential energy difference in the moduli sector
between the minima during and after thermal is small, so the entropy
production is not so large as to cause cosmological problems. If there
is more than one bout of thermal inflation, the moduli problem is
solved for a wide range of the values of $\Sigma_i$
\cite{thermalinfl}. Given these successes, one might even argue that
flat directions leading to thermal inflation should be a {\em
necessary} ingredient of cosmologically viable supergravity models; in
turn this motivates the possibility of structure in the spectrum of
the density perturbation generated during primordial inflation.

To summarize, intermediate-scale symmetry breaking can lead to two
eras of multiple inflation. The first era (primordial inflation)
occurs at a high scale with a Hubble parameter as high as
$\sim10^{10}$~GeV during which the density perturbation responsible
for the large-scale structure we observe today must have been
produced. The second era (thermal inflation) occurs at a lower scale
with a Hubble parameter as low as $\sim10^{-6}$~GeV and must last
$\lesssim25-30$ e-folds if the density perturbation produced during
the primordial inflation era is not to be erased. Each stage of
primordial inflation may have a corresponding stage of thermal
inflation but this is model dependent, being sensitive to the value of
the reheat temperature following primordial inflation.

During the primordial inflationary era, each burst of inflation lasts
about 10--20 e-folds. The density perturbation produced during each
burst is nearly scale-invariant but with differing amplitudes. Between
each burst there is a brief period during which scale-invariance is
badly broken and the spectral index departs from unity
(``notches''). The bursts of thermal inflation last about
$\ln(\Sigma/m_W)/2\approx10-15$ e-folds, for
$\Sigma\sim(10^{-8}-10^{-4})M$. Baryogenesis, and subsequently
nucleosynthesis, must of course occur afterwards.

Clearly the picture of inflation we have drawn is much more complex
than is normally assumed. We now consider possible tests of this
picture through astronomical observations.

\section{Observational Tests}

Observations of large-scale structure in the universe are usually
expressed in terms of the power spectrum $P(k)$ of galaxy
clustering. Given this we must first obtain the {\em linear} spectrum
of matter fluctuations by allowing for the effects of non-linear
evolution as the density contrast becomes of order
unity.\footnote{\footnotesize The distribution of galaxies may be
 `biased' with respect to the distribution of the (dark) matter in
 being clustered to a greater or lesser extent. To avoid such
 complications in normalization of the power spectrum, we focus
 therefore on the {\em spectral index} since such biasing cannot be
 strongly scale-dependent \cite {lss,peacock}.} Then one can
deconvolute the linear transfer function $T(k)$, appropriate to the
(dark) matter content, which tracks the scale-dependent rate of growth
of linear perturbations. Recall that the spectrum of rms mass
fluctuations after matter domination (per unit logarithmic interval of
wavenumber $k$) is \cite{lss}
\begin{equation}
 \Delta^2 (k) \equiv \frac{k^3 P(k)}{2\pi^2} 
              = \delta^2_{\rm H}(k)\ T^2 (k) \left(\frac{k}{aH}\right)^4\ .
\end{equation}
For CDM we use the usual parametrisation \cite{lss}, 
\begin{equation}
T (k) = \left[1 + \left\{a k + (b k)^{3/2} + (c k)^2
\right\}^{\nu}\right]^{-1/\nu}\ ,
\label{Tk}
\end{equation}
with $a=6.4\Gamma^{-1}h^{-1}{\rm Mpc}$, $b=3\Gamma^{-1}h^{-1}{\rm
Mpc}$, $c=1.7\Gamma^{-1}h^{-1}{\rm Mpc}$ and $\nu=1.13$, where the
`shape parameter' is $\Gamma\simeq\Omega{h}\,{\rm e}^{-2\Omega_{\rm
N}} $\cite{pd94}, $\Omega_{\rm N}$ being the fraction of the
critical density in nucleonic matter .

The standard tool for studying non-linear evolution is a N-body
computer simulation which directly solves the governing equations
\cite{nbody}.  However such simulations become numerically expensive
as the number of particles or the dynamic range is increased and a
more affordable approach is to identify a mapping between the linear
(L) and nonlinear (NL) spectra \cite{pd94}:
\begin{equation}
 \Delta^2_{\rm NL}(k_{\rm NL}) = f_{\rm NL} (\Delta^2_{\rm L} (k_{\rm L}))\ ,
 \qquad k_{\rm L} = [1 + \Delta^2_{\rm NL} (k_{\rm NL})]^{-1/3} k_{\rm NL}\ .
\end{equation}
Here the scaling ansatz \cite{hklm} relates $\Delta^2_{\rm NL}$ at a
given scale to $\Delta^2_{\rm L}$ at a {\em different} scale, taking
into account the shrinking (in comoving coordinates) of nonlinear
density fluctuations during the nonlinear evolution. A simple physical
model of clustering processes predicts a similar correlation between
the linear and nonlinear power spectra with a piecewise scaling
function corresponding to the linear, quasilinear (dominated by
scale-invariant radial infall) and nonlinear (dominated by nonradial
motions and mergers) regimes \cite{paddy}.

The universal scaling function, $f_{\rm NL}$ is obtained by a
multiparameter fit to N-body simulations \cite{nbody}. Two suggested
forms are
\begin{equation}
 \mbox{JMW \cite{jmw}:} \quad f_{\rm NL}(x) = 
  x \left[\frac{1 + ay + by^{2} + cy^{3} + dy^{3.5} + ey^{4}} 
  {1 + fy^{3}}\right]^{1/2} , \quad
  y = \frac{x}{[(3 + n)/3]^{1.3}}\ ,
\label{nonlinfjmw}
\end{equation}
with $a=0.6$, $b=1.0$, $c=-0.2$, $d=-1.5$, $e=1.0$ and $f=0.0037$, and 
\begin{equation}
\mbox{PD \cite{pd96}: } \quad f_{\rm NL} (x) = 
  x \left[\frac{1 + B \beta x + (A x)^{\alpha\beta}}{1 + \{(A x)^\alpha g^3
  (\Omega)/(Vx^{1/2})\}^\beta}\right]^{1/\beta} ,
\end{equation}
where $A=0.482\,(1+n/3)^{-0.947}$, $B=0.226\,(1+n/3)^{-1.778}$, $
\alpha=3.310\,(1+n/3)^{-0.244}$, $\beta=0.862\,(1+n/3)^{-0.287}$, $
V=11.55\,(1+n/3)^{-0.423}$. The former expression (\ref{nonlinfjmw})
has been demonstrated \cite{bg} to work well for the APM data while
the latter expression improves on a similar form given earlier
\cite{pd94} for general CDM spectra.

\begin{figure}[tbh]
\center{\epsfxsize5.5in\epsffile{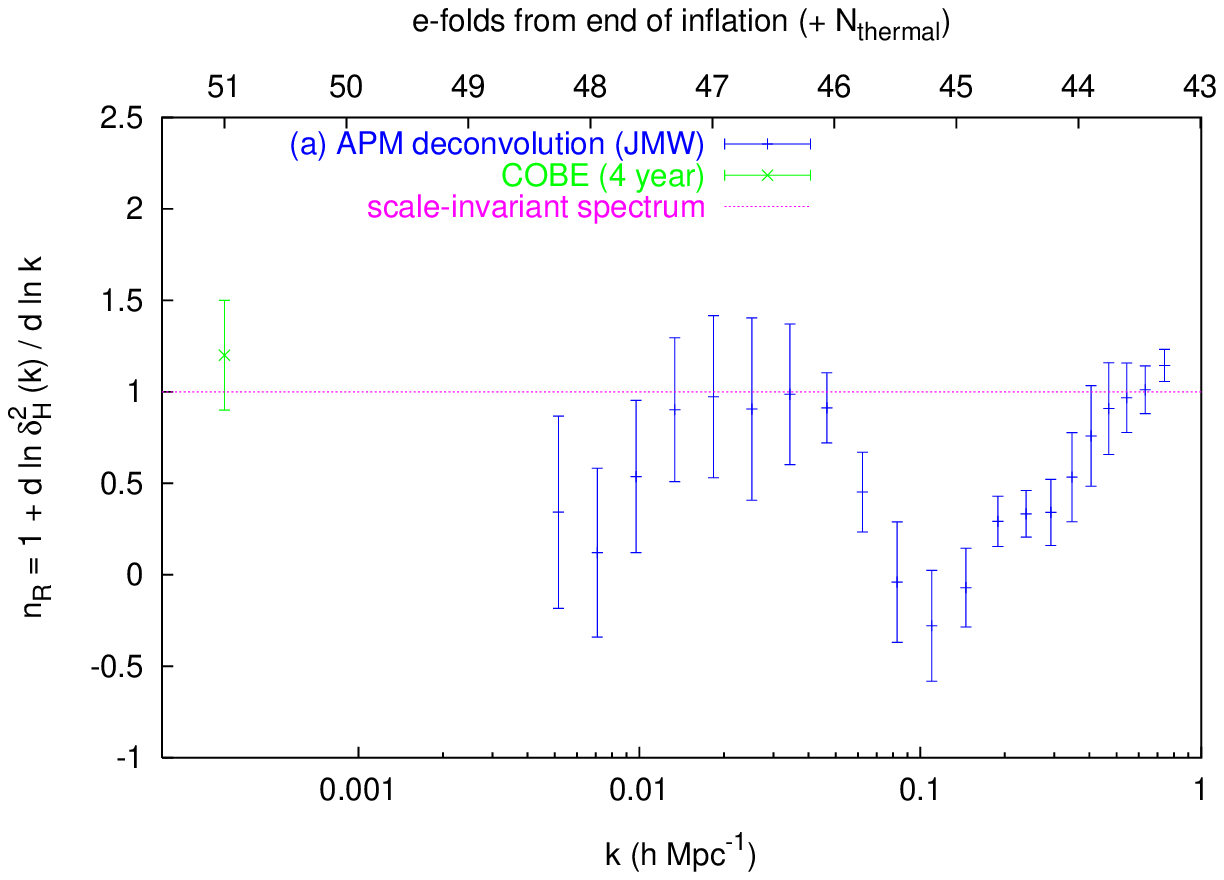}} 
\center{\epsfxsize5.5in\epsffile{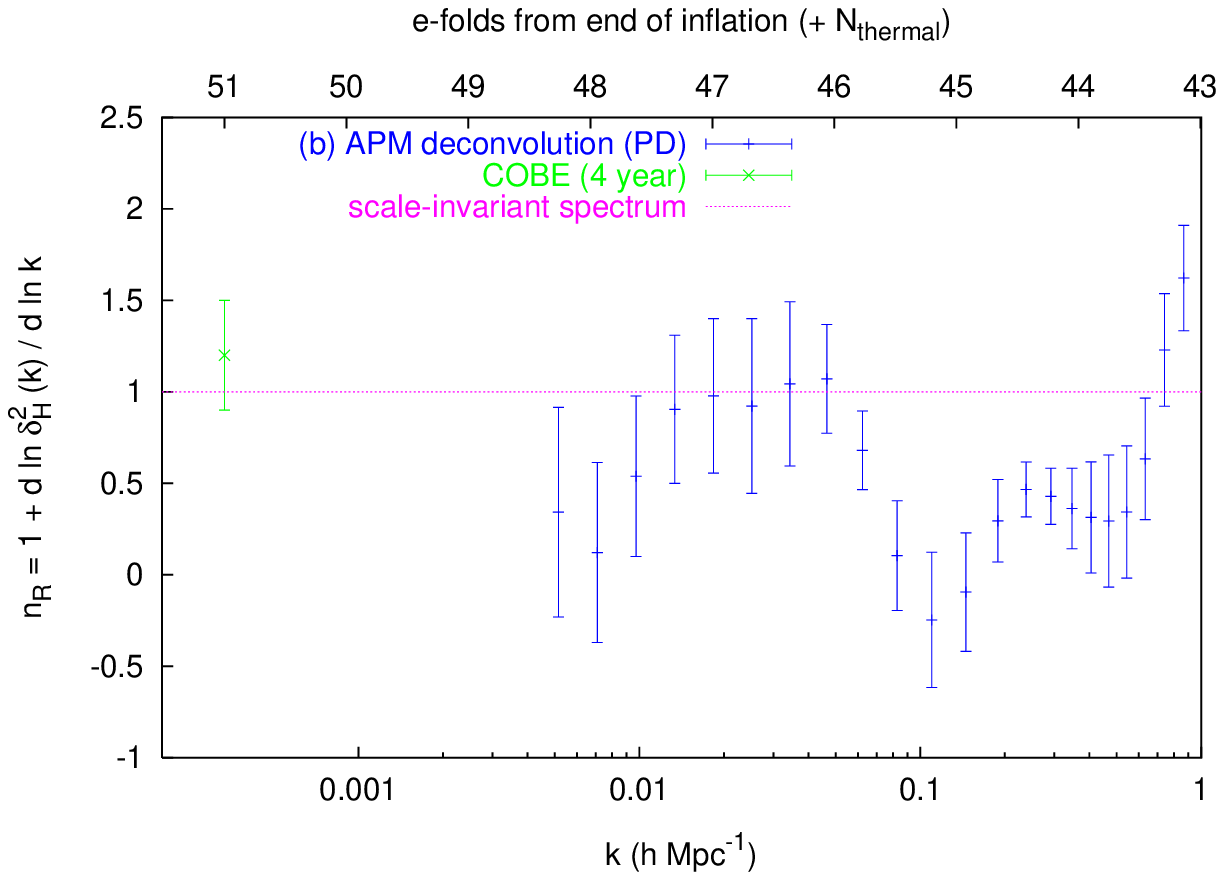}} 
\bigskip
\caption{The spectral index of the primordial density perturbation
implied by the APM data, assuming a critical CDM universe with
$\Omega_{\rm N}=0.05 $, $h=0.5$, and using two suggested procedures to
extract the linear spectrum. A scale-invariant spectrum, the usually
assumed generic prediction of inflation, is shown for comparison. The
``notch'' seen at $k\sim0.1h$~Mpc$^{-1}$ is in accordance with the
spectral feature expected due to an intermediate scale symmetry
breaking phase transition.}
\label{nk}
\end{figure}

Using these expressions, we have reconstructed the linear power
spectrum from the three-dimensional $P(k)$ inferred from the angular
correlation function of galaxies in the APM survey \cite{apm}. This
data set is particularly valuable as it is not subject to
redshift-space distortion effects and currently provides the most
accurate and deepest probe of galaxy clustering in the universe. We
determine the spectral index $n(k)\equiv{\rm d}\ln\,P(k)/{\rm
d}\ln\,k$ as a function of $k$, so that the {\em primordial} spectral
index $n_{\cal R}(k)$ (\ref{n}) can be obtained simply by
subtracting off the slope of the transfer function (\ref{Tk}) for each
value of $k$. This is shown in figure~\ref{nk} along with the data
point $n_{\cal R}=1.2\pm0.3$ at $k_{\rm
COBE}^{-1}\sim\,H_0^{-1}\simeq\,3000\,h^{-1}$\,Mpc obtained from the
4-year COBE data \cite{cobe}. The number of e-folds before the end of
inflation when this particular scale crossed the Hubble radius is
usually obtained from the formula
\begin{eqnarray}
 N_{\rm COBE} \equiv N_\star(k=k_{\rm COBE}) & \simeq 51 
  & + \ln\left(\frac{k^{-1}}{3000h^{-1}\,{\rm Mpc}}\right) 
    + \ln\left(\frac{V_\star}{3\times10^{14}\,{\rm GeV}}\right) 
    + \ln\left(\frac{V_\star}{V_{\rm end}}\right) \nonumber \\
 && - \frac{1}{3}\ln\left(\frac{V_{\rm end}}{3\times10^{14}\,{\rm GeV}}\right) 
    + \frac{1}{3}\ln\left(\frac{T_{\rm reheat}}{10^5\,{\rm GeV}}\right),
\label{Nstar} 
\end{eqnarray}
where we have inserted appropriate numerical values for the various
energy scales in our primordial inflation model
\cite{success,natural}. This formula is derived assuming that the
thermal history is standard following reheating after primordial
inflation. However as we have seen there will be subsequent periods of
thermal inflation, each of which will reduce $N_{\rm COBE}$ further by
$\ln(\gamma)/3\approx15-20$ e-folds, where $\gamma$ is the factor by
which the comoving entropy is increased. (Obviously only 1--2 such
periods are acceptable.) Denoting the total correction necessary by
$N_{\rm therm}\equiv\Sigma_i{\cal N}^i_{\rm therm}$, we write
\begin{equation}
 N_\star (k) \simeq N_{\rm COBE} - N_{\rm therm}\ ,
\end{equation}
which is marked on the top axis in figure~\ref{nk}.

It is seen in figure~\ref{nk} that there is indeed a distinct
``notch'' in the primordial spectrum at around
$k\sim0.1h$~Mpc$^{-1}$. The slope drops sharply from about unity to a
small negative value before rising again towards unity, all within
about 2 e-folds. Note that this feature is robust with regard to which
analytic form is used to recover the linear spectrum from the APM
data. A similar feature is also evident in the data from the IRAS
survey \cite{peacock}. Given that this feature appears at a spatial
scale of order the horizon at the epoch of (dark) matter domination,
it is perhaps natural to interpret it as reflecting a possible
departure from the pure cold dark matter paradigm. However we pursue
the alternative possibility, {\em viz.} that it reflects a departure
from slow-roll inflation and is present in the {\em primordial}
fluctuation spectrum, not just in the present density perturbation. to
discriminate between these possibilities requires better data such as
would be provided by the ongoing 2DF redshift survey. Note that a
second ``notch'' may be present at around $k\sim0.01h$~Mpc$^{-1}$
although to substantiate this would require data at scales in between
those probed by COBE and by galaxy surveys, such as would be provided
by intermediate scale CMB anisotropy experiments.

\begin{figure}[tbh]
\center{\epsfxsize5.5in\epsffile{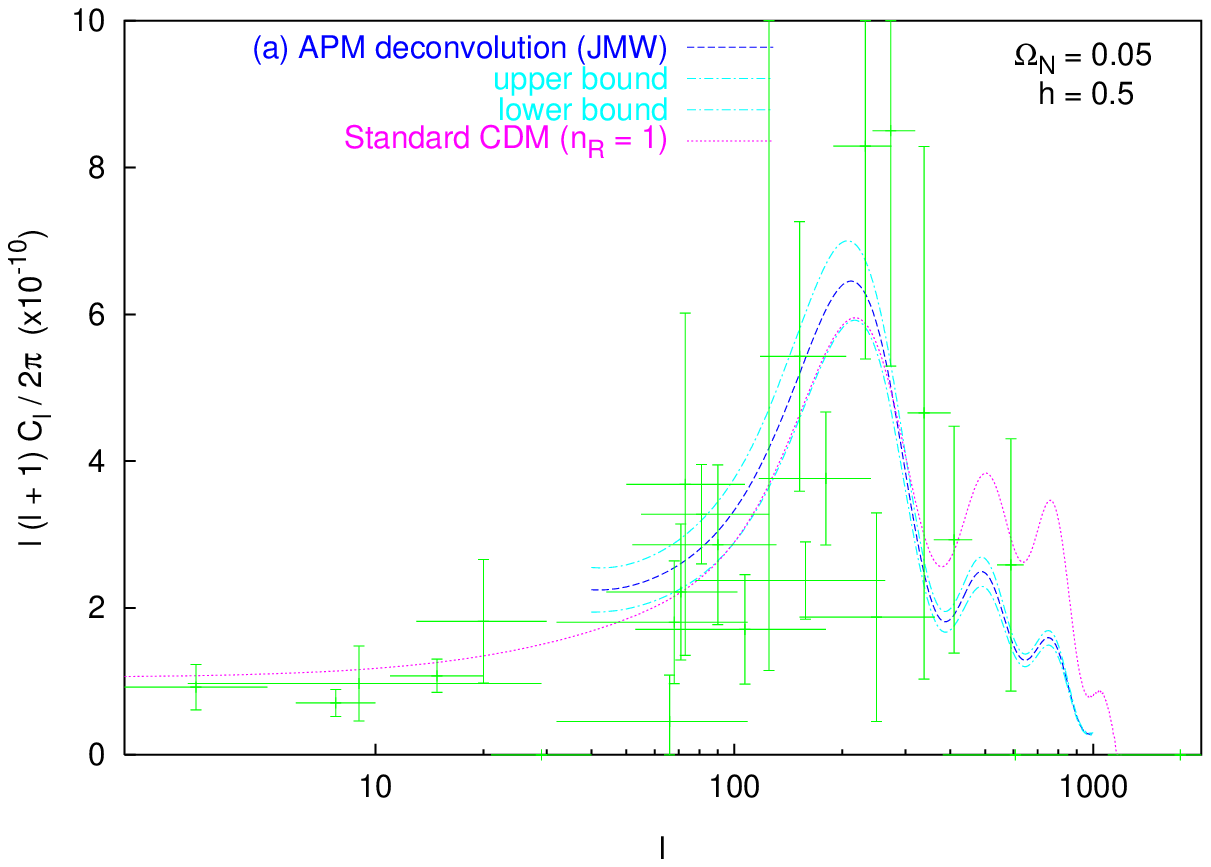}} 
\center{\epsfxsize5.5in\epsffile{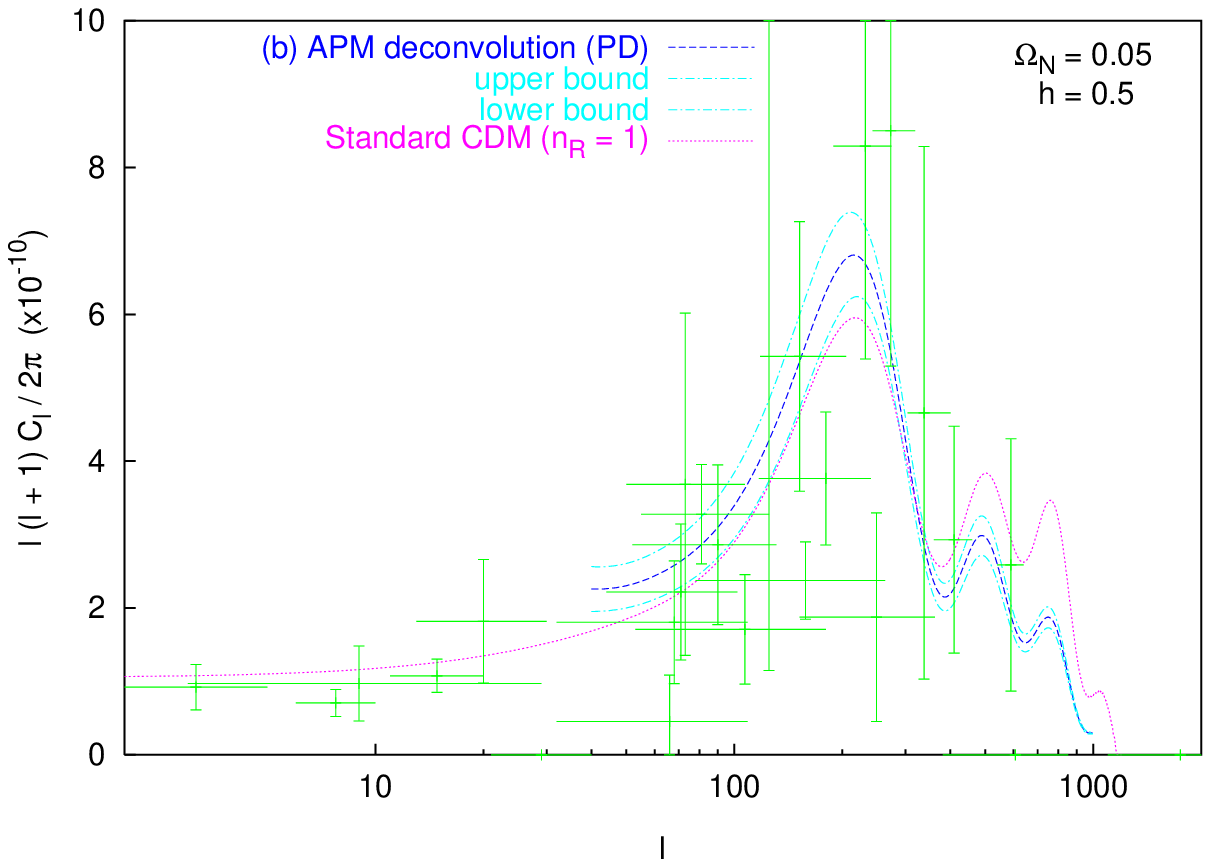}} 
\bigskip
\caption{The power spectrum of CMB anisotropy implied by the APM data,
assuming a critical CDM universe with $\Omega_{\rm N}=0.05$, $h=0.5$,
and using two suggested procedures to extract the linear spectrum. The
upper and lower envelopes correspond to the quoted $\pm1\sigma$ errors
in the APM data averaged over 4 zones. The expectation for a
COBE-normalized standard CDM model (with $n_{\cal R}=1$) is shown for
comparison, along with an unbiased compendium of observational data.}
\label{cl}
\end{figure}

It is interesting to ask what would be the expectation for the angular
anisotropy of the CMB if the primordial spectrum indeed has the shape
shown in figure~\ref{nk}. To determine this we have run the COSMICS
code \cite{bert}, which numerically solves the coupled linearized
Boltzmann, Einstein and fluid equations for the perturbation in the
photon phase space distribution, with the primordial spectrum
reconstructed from the APM data as the input. (We have forced the APM
slope to be unity at very large scales for consistency with COBE; the
implied quadrupole moment is however higher than the COBE measurement
by a factor of about $1.5$ reflecting the bias of APM galaxies with
respect to the dark matter.) The multipoles are plotted in
figure~\ref{cl}, taking $\Omega_{\rm N}=0.05$, $h=0.5$, along with a
compendium of recent observational data \cite{cmbobs}. We see that the
heights of the secondary Doppler peaks are suppressed relative to the
prediction of the COBE-normalized standard CDM model, which assumes
$n_{\cal R}=1$. (This conclusion would be further strengthened if the
the APM curve were to be lowered by a factor of 1.5 in order to be
normalized to COBE.)  Again this prediction is robust regardless of
the analytic form is used to recover the linear spectrum from the APM
data. A precision measurement of this region of the CMB angular power
spectrum will be made by the forthcoming PLANCK (formerly
COBRAS/SAMBA) mission.

If the picture we have presented receives observational verification
it will be the first step in determining the structure of the physical
theory at high energies from astronomical measurements of the density
perturbation. Other interesting possibilities are raised by multiple
inflation. Since inflation now lasts for a relatively short period,
the density parameter $\Omega$ may be close to but not quite equal to
unity \cite{open}, depending on exactly how many e-folds of inflation
the universe has undergone. Regarding structure formation, given that
more than one scalar field is involved, the possibility of generating
non-gaussian or even isocurvature fluctuations arises
\cite{toy,isocurv}. The latter is particularly relevant to the
possibility that the dark matter consists of axions \cite{axion}. It
has been noted that thermal inflation can relax the cosmological upper
bound on the axion scale by several orders of magnitude
\cite{crisis,thermalinfl}, so that string-theoretic axions may well
constitute the dark matter \cite{stringaxion}. Alternatively, the
relic abundance of bound states from the hidden sector (`cryptons')
may be diluted sufficiently so as to enable them to constitute the
dark matter \cite{crypton}; being metastable, such particles may even
be the source of the observed very high energy cosmic rays \cite{prep}.

\vfill \noindent {\bf Acknowledgments.} We are grateful to Enrique
Gaztan\~{a}ga, for providing the APM data and for very helpful
correspondance concerning the linear spectrum. We thank David Lyth for
motivating us to examine the effects of intermediate scales and for
stimulating discussions, as well as George Efstathiou and John Peacock
for helpful comments.

\end{document}